\documentclass[twocolumn,showpacs,preprintnumbers,amsmath,amssymb,superscriptaddress]{revtex4-1}

\usepackage{graphicx,color,amssymb,amsthm,hyperref}
\usepackage{dcolumn}
\usepackage{bm}
\usepackage{bbm}

\newcommand{\ket}[1]{|#1\rangle}

\usepackage{multirow,array,rotating}
\newcolumntype{M}{>{$\vcenter\bgroup\hbox\bgroup}c<{\egroup\egroup$}}
\newcommand{\ve}{\boldsymbol}

\def\>{\rangle}
\def\<{\langle}

\def\opt{ {\mathrm{opt}} }

\DeclareMathOperator{\Tr}{Tr}

\begin{document}


\title{Single-qubit thermometry}

\author{Sania Jevtic}
\affiliation{Mathematical Sciences, John Crank 501, Brunel University, Uxbridge UB8 3PH, United Kingdom}
\author{David Newman}
 \affiliation{HH Wills Laboratory, Bristol University, Tyndall Avenue, Bristol BS8 1TL}
\author{Terry Rudolph}
 \affiliation{Department of Physics, Imperial College London, London SW7 2AZ, United Kingdom}
\author{T.\ M.\ Stace}
\affiliation{ARC Centre for Engineered Quantum Systems, University of Queensland, Brisbane 4072, Australia}

\date{\today}

\begin{abstract}
Distinguishing hot from cold is the most primitive form of thermometry.  Here we consider how well this task can be performed using a single qubit to distinguish between two different temperatures of a bosonic bath.  In this simple setting, we find that letting the qubit equilibrate with the bath is not optimal, and depending on the interaction time it may be advantageous for the qubit to start in a state with some quantum coherence. We also briefly consider the case that the qubit is initially entangled with a second qubit that is not put into contact with the bath, and show that entanglement allows for even better thermometry.

\end{abstract}

\pacs{03.67.-a 03.65.Ta 03.65.Yz}
\maketitle

A standard classical thermometer begins in thermal equilibrium and is used by observing a subsequent change in its macroscopic state, once the thermometer has equilibrated with the object (henceforth simply called the bath) whose temperature it indicates. A good thermometer is small, so that it does not significantly disturb the temperature of the bath in the process, although the number of microstates consistent with its initial and final macrostates will still be large. The (weak) coupling between the thermometer and bath leads to quasi-static evolution \cite{Blundell}.

Here we consider a quantum extreme of temperature measurement that differs from this classical scenario in several ways. Firstly, we will be interested in the smallest possible thermometer, namely a single qubit. This is motivated by the observation that for nanoscale experiments, the bath (e.g. a micromechanical resonator) may itself be very small, necessitating an even smaller thermometer (a motivation also for \cite{Brunelli1}). Secondly, we will be interested in temperature measurements that take place over times potentially much shorter than the time it takes for the qubit thermometer to get close to equilibrium with the bath, as investigated in \cite{Stace}. This is motivated by the observation that certain baths (e.g. a small cloud of cold atomic gas) may be difficult to create and then maintain for long times \cite{McKay}. Two-level atomic quantum dots have previously been suggested as thermometers for BECs \cite{Bruderer,Sabin}, but there are fundamental limits to how precisely we can measure the temperature of quantum gases \cite{Marzolino}. Finally the initial state of the qubit thermometer need not be thermal, and therefore at intermediate times the state of the qubit also need not be close to thermal. This will allow us to examine whether quantum coherence can play a useful role in such thermometry.

We consider an extremely simple version of thermometry, where the goal is simply to determine whether a standard bosonic bath is cold or hot, i.e. at temperature $T_1$ or temperature $T_2>T_1$ where the two possible bath temperatures $T_1$, $T_2$ are known. We focus on this simple setting because it is sufficient for revealing curious physical principles. Our goal is not to propose a practical implementation of thermometry, instead we want to strip away many of the complex and technical choices that one generally must make in quantum metrology. By abstracting and simplifying the thermometry scenario we ascertain the importance of transient dynamics. Note that qubit thermometry for an unknown bath temperature is analysed in \cite{Sanpera_Qubit_Thermo}. Generalizations to non-Markovian baths are considered in \cite{Brunelli2,Higgins}.

We use units where Planck's constant $\hbar$ and Boltzmann's constant $k_B$ are both 1, and time $\tau$ appears in the ``dimensionless time'' $ t = \gamma \tau$, where 
 $\gamma$ is the qubit's spontaneous emission rate induced by the coupling to the bath. By convention we take the ground state of the qubit to be $\ket{1}$ and the excited state to be $\ket{0}$. The energy difference $\omega$ of these states must be large compared to $\gamma$ for a standard Markovian decoherence model (master equation) to apply \cite{Carmichael,Nielsenchuang}, which it will do on timescales $\tau \gg1/\omega,1/T$. The bath temperatures $T_i$ must also be large compared to $\gamma$, however the average boson occupation number for the bath $\bar{N}_i=(e^{\omega/T_i} - 1)^{-1}$ can vary arbitrarily, as the ratio $\omega/T_i$ is not constrained in order for the analysis to apply.

We work with a model in which the qubit thermometer interacts with a bath, and the bath is
in thermal equilibrium. 
The qubit, which is represented by a two-level system, is prepared in a known state with Bloch vector $\ve r(0)=(r_x,r_y,r_z)=(R\sin\theta\cos\phi,R\sin\theta\sin\phi,R\cos\theta)$. The qubit is coupled to the bath at time $t=0$ and they interact for a fixed time $t$ at which point the qubit is decoupled from the bath and a measurement is made. During the interaction, the qubit's state has evolved according to a standard master equation \cite{Carmichael}. At time $t$ its Bloch vector is given by
\begin{equation}
\ve r(t,T) =
\left(
  \begin{array}{c}
    r_x e^{-(1+2\bar{N}) t/2} \\
    r_y e^{-(1+2\bar{N}) t/2} \\
    \frac{e^{-(1+2\bar{N})t}\left(1+(1+2\bar{N})r_z \right)-1}{1+2\bar{N}} \\
  \end{array}
\right)
\end{equation}
The mean bosonic occupation number $\bar{N}$ always appears as
\begin{eqnarray}
1+2\bar{N} &=& \coth\left(\frac{1}{2T}\right):= n(T)
\end{eqnarray}
and we denote $n(T_i)=n_i$, $\ve r(t,T_i)=\ve r_i(t)$, $i=1,2$.
Given that we know that at time $t$ the qubit is either in $\ve r_1(t)$ or $\ve r_2(t)$, we measure the qubit in the basis that has the highest probability to distinguish the two possible states. For two qubits, the maximal probability is
 $$\frac 1 2 \left(1 + \frac 1 2 |\ve r_{1}(t) - \ve r_{2}(t)|\right)=:\frac 1 2 \left(1 + \frac 1 2 \Delta(\ve r_{1}(t),\ve r_{2}(t))\right), $$ 
 where $|...|$ denotes the Euclidean distance between the two vectors. We find
\begin{eqnarray}\label{delta}
\Delta(\ve r_{1}(t),\ve r_{2}(t))^2 &=A^2R^2 \sin^2\theta + \left(BR\cos\theta-C\right)^2, \nonumber
\end{eqnarray}
where
\begin{eqnarray}\label{delta}
A=A(n_1,n_2,t)&:=&e^{-n_1t/2} - e^{-n_2t/2}\nonumber \\
B=B(n_1,n_2,t)&:=&e^{-n_1t} - e^{-n_2t} \nonumber \\
C=C(n_1,n_2,t)&:=&\frac{1-e^{-n_1 t}}{n_1} - \frac{1-e^{-n_2 t}}{n_2}\nonumber
\end{eqnarray}
Note any $\phi$ dependence in the initial state drops out. Since $A,B,C\ge 0$ for $T_1<T_2$ (implying $n_1<n_2$), our first observation is that $|\ve r_1(t) - \ve r_2(t)|$ will be maximized for $\pi/2 \leq \theta \leq \pi$, and within this range we should take $R=1$, i.e. the initial state of the qubit to be pure. This already tells us that, for example, starting the qubit in the excited state $\theta=0$ will not be optimal. However, because there are interesting dynamics for states with $0\leq \theta \leq \pi/2$ we will occasionally consider $\theta$ over its full range.
Henceforth we always take the initial state pure, and therefore
\[
\Delta(t,\theta|T_1,T_2)^2=A^2\sin^2\theta  + \left(B\cos\theta -C\right)^2.
\]
For long times the qubit will reach equilibrium, the distinguishability of the two equilibrium states is determined by $\Delta(\infty,\theta|T_1,T_2):=\Delta_{\infty}=n_1^{-1}-n_2^{-1}$. 

\begin{figure}[h!]
\includegraphics[width=\columnwidth]{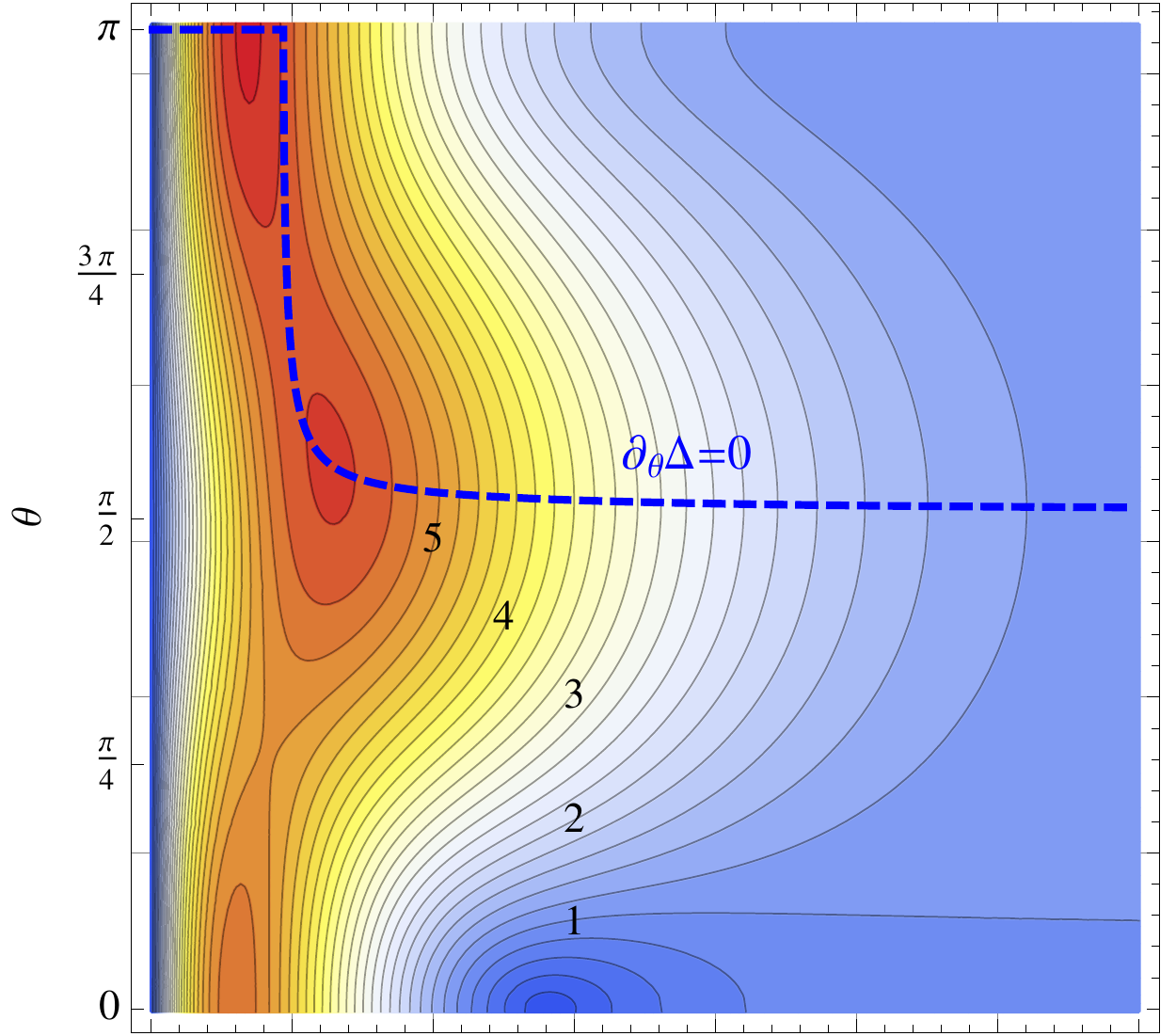}
\includegraphics[width=\columnwidth]{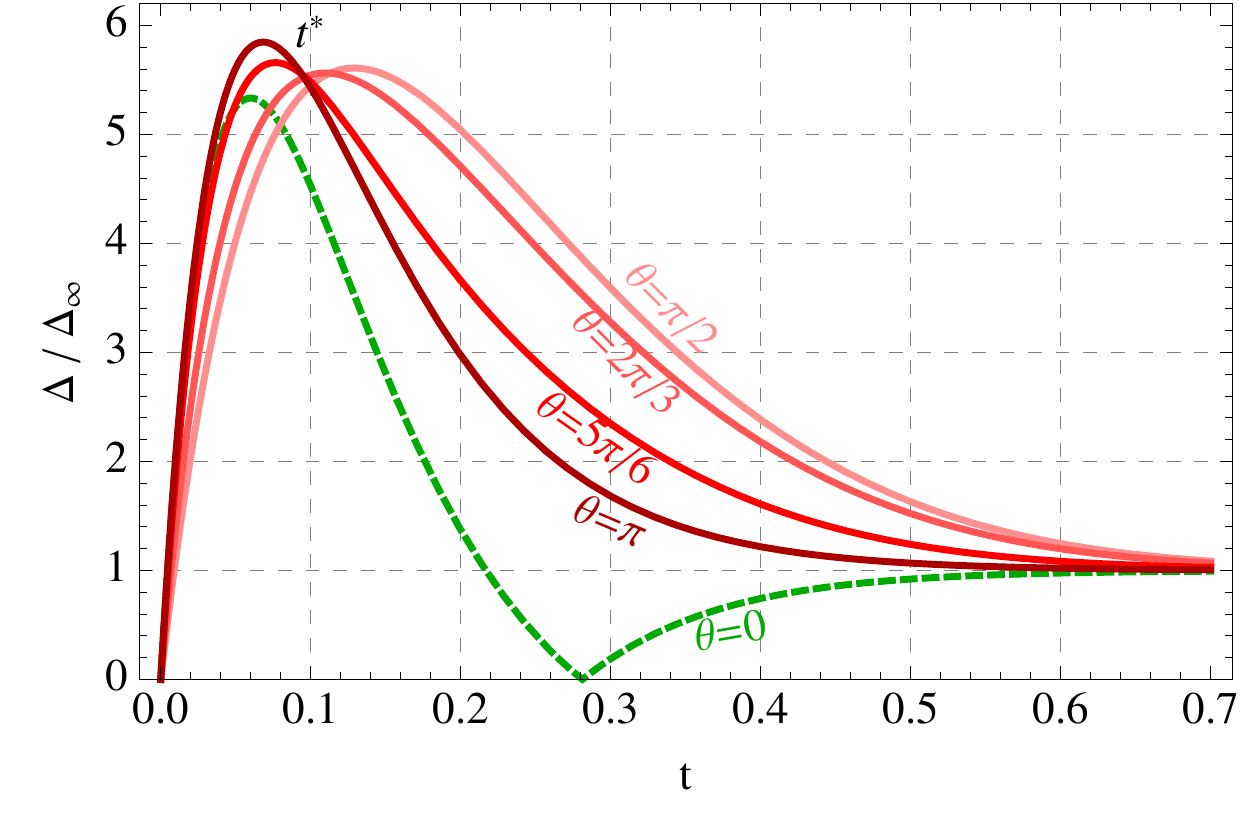}
\caption{(Color online) Variation of the Euclidean distance $\Delta$, normalised to $\Delta_{\infty}$, versus $t$ (dimensionless) and $\theta$, for $n_1$=12, $n_2=20$. (Top) Contours of $\Delta(t,\theta|T_1,T_2)/\Delta_\infty$.  The blue (dashed) line indicates the value of $\theta$ that maximises $\Delta$ for each $t$. (Bottom)
$\Delta/\Delta_\infty$ for values of various $\theta$.}
\label{fig1}
\end{figure}

To quantify the performance of the qubit thermometer compared to that of its thermalized counterpart, we normalise the Euclidean distance $\Delta(t,\theta|T_1,T_2)$ to $\Delta_{\infty}$.  Figure \ref{fig1} illustrates the performance of the qubit thermometer for a particular combination of possible bath temperatures (corresponding to $n_1=12, n_2=20$).  From the figure we immediately note several interesting features. These features hold for all finite temperatures, unless stated otherwise.
\begin{itemize}
\item It is certainly possible to do better than waiting until the qubit `thermometer' equilibrates with the bath, i.e. at some intermediate times the two possible trajectories are further apart than the equilibrium distance.
\item The global maximum in the distinguishability is a sharp peak after a short time, corresponding to the case of initialising the qubit in the ground state ($\theta=\pi$). Therefore if the interaction time can be finely tuned this strategy is optimal. 
\item After some finite time the optimal initial state rapidly changes from being the ground state to something close to maximally coherent, $\theta=\pi/2$. Note that the optimal initial coherent state depends on the choice of $n_i$.
\item At each instant of time, the optimal strategy has $\theta\in[\pi/2,\pi]$.
\item Initial states that begin with coherence decay to the equilibrium distance $\Delta_{\infty}$ slower than those without.
\item Beginning in the excited state $\theta=0$ there is a finite time where the trace distance goes to 0.
\end{itemize}
We now turn to elucidating a few of these observations.

For what fixed time of interaction $t$ does using coherence in the initial state become better than starting the qubit in the ground state? Intriguingly we find that there is a fixed finite time $t^*$ up to which starting in the ground state is optimal, and after which starting in a state with some coherence performs better. It can be readily shown that $t^*$ is the solution to $A^2 - B^2 = BC$ for fixed $n_1,n_2$, although there appears to be no closed form solution in this case.   

For times $t<t^*$, the optimum strategy (which is also globally optimal) is to initialise the qubit in the ground state, for which the distinguishability 
\begin{eqnarray}
\Delta_g = B+C.
\end{eqnarray}
This reaches a maximum at time
\begin{equation}
t_g=\frac{1}{n_2-n_1}\ln\left( \frac{n_2-1}{n_1-1}\right)=\frac{1}{2(\bar{N}_2-\bar{N}_1)}\ln\left(\frac{\bar{N}_2}{\bar{N}_1}\right).
\end{equation}
This is the optimum time to measure the qubit for maximal distinguishability. In the singular case that $n_1=1$, corresponding to the cold bath being in its ground state, then no maximum is achieved since $\Delta_g$ increases monotonically to the equilibrium value of $1-n_2^{-1}$.

For times $t>t^*$ the optimal value of $\theta$ is given by
\begin{equation}
\theta^{\opt} = \pi-\arccos(BC/(A^2-B^2)).
\end{equation}
For this choice of initial state the maximum square distance is
\begin{equation}\Delta_{opt}^2=A^2\left(1+\frac{C^2}{A^2-B^2}\right).\end{equation} We have therefore obtained a complete solution to the basic problem, and we now turn our attention to certain special cases.

For fixed $n_1,n_2$ and $t>t^*$ the value of $BC/(B^2-C^2)$ rises rapidly as a function of time to an asymptotic value of $n_1^{-1}-n_2^{-1} = \Delta_\infty$. (The rise is typically but not always monotonic - for certain parameter choices it can actually slightly exceed this value then come back down to it). Therefore for $n_1,n_2$ moderately large or $n_1 \approx n_2$ so that $\Delta_\infty \approx 0$, the optimal choice of state for $t>t^*$ will be approximately the maximally coherent state $\theta=\pi/2$. Otherwise, except within a very short time interval, it will be some coherent state with angle $ \pi-\arccos(\Delta_\infty)$.

In the special case of maximal coherence in the initial state of the qubit, i.e. $\theta=\pi/2$ we have
\begin{equation}
\Delta_c^2 =  A^2+C^2.
\end{equation}
It is not simple to find the time at which this is maximized. However, for moderately large values of $n_1,n_2$ and over the short times we are interested in, the $A^2$ term dominates the $C^2$ term and $\Delta_c$ reaches its maximum at approximately the time that $A$ has a maximum, which is
\begin{equation}
t_{c}\approx \frac{2}{n_2-n_1}\ln\left( \frac{n_2}{n_1}\right).
\end{equation}
This is roughly twice the time at which the corresponding expression for starting in the ground state reaches its maximum.
Alternatively, we can compare the ground and maximally coherent state distances for short times:
\begin{eqnarray}
\Delta_g &\approx& (n_2-n_1) t \\
\Delta_c &\approx& (n_2-n_1)\frac{t}{2},
\end{eqnarray}
indicating the ground state initial condition increases towards its maximum about twice as fast as the maximally coherent state.

To obtain some feel for how the coherent trajectories make for more robust thermometry at longer times, consider the simple special case of $n_2=2n_1$, for which, at long times, we have:
\begin{eqnarray}
\Delta_g &\approx& \frac{1}{2n_1}+\left(1-\frac{1}{n_1}\right)e^{-n_1 t}\\
\Delta_c &\approx& \frac{1}{2n_1}+\left(n_1-\frac{1}{n_1}\right)e^{-n_1 t}.
\end{eqnarray}
Thus if the colder bath is actually quite warm, the coherent trajectories last longer.

An excited state probe $\theta = 0$ has the feature that the paths the qubit follows under $T_1, T_2$ cross at some finite time, leading to the cusp-shaped curve (dashed) in Fig.\ 1(bottom), which occurs when $B=C$. This occurs basically because the trajectory $\ve r_2(t)$ initially moves faster down the z-axis of the Bloch sphere (going roughly as $e^{-n_2 t}$) but near the equilibrium point it slows higher up that axis (since it is hotter). As such, there is a finite time where the colder trajectory $\ve r_1(t)$ catches up and the two cross.

We now consider briefly the case where we still send only a single qubit through the bath for a finite time, but it is initially entangled with a second qubit. After the interaction the optimal joint measurement on both qubits is performed. We expect a pure entangled state to enhance the distinguishability: the higher-dimensional system has a larger Hilbert space to explore with double the number of orthogonal, perfectly distinguishable states. This has a favourable effect on the state discrimination because of the increased capacity with which to encode information about the bath.

From numerical study it appears that the optimal state is not necessarily maximally entangled. This is perhaps not surprising: Fujiwara \cite{Fujiwara} concludes that entanglement deteriorates the information contained in the outputs of a generalised amplitude damping channel. In his study, a state of the form $\sqrt{1-\alpha}\ket{01} - \sqrt{\alpha}\ket{10}$, with $\alpha \neq \frac 12$ in general, maximises the information about the rate of dissipation (as measured by the ``symmetric logarithmic derivative Fisher information''). Temperature may simply be viewed as another parameter, and so finding that the optimal thermometer is not maximally entangled fits in with previous observation.

Let us consider the maximally entangled case since it is analytically tractable.
If the two qubits $a$ and $b$ are initially prepared in the maximally entangled state $\ket{\phi^+}=(\ket{00}+\ket{11})/\sqrt 2$, after qubit $a$ has interacted with the bath for time $t$ we find that 
\begin{multline}
\rho_{ab}(n_1,t)-\rho_{ab}(n_2,t)=\\
                                   \frac{1}{4}\left(
                                    \begin{array}{cccc}
                                      B+C & 0& 0 & 2A \\
                                      0 & -(B+C) & 0 & 0 \\
                                      0 & 0 & C-B & 0 \\
                                      2A & 0 & 0 & -(C-B) \\
                                    \end{array}
                                  \right).\\
\end{multline}
The trace distance is then proportional to
\begin{multline}
\Delta_{\phi^+}:=\Tr\left|\rho_{ab}(n_1,t)-\rho_{ab}(n_2,t)\right| \\
=\frac{1}{4}\Big(|B+C|+|B-C|\\+|B+\sqrt{4A^2+C^2}|+|B-\sqrt{4A^2+C^2}|\Big)
\end{multline}

\begin{figure}[t!]
\includegraphics[width=3.4in]{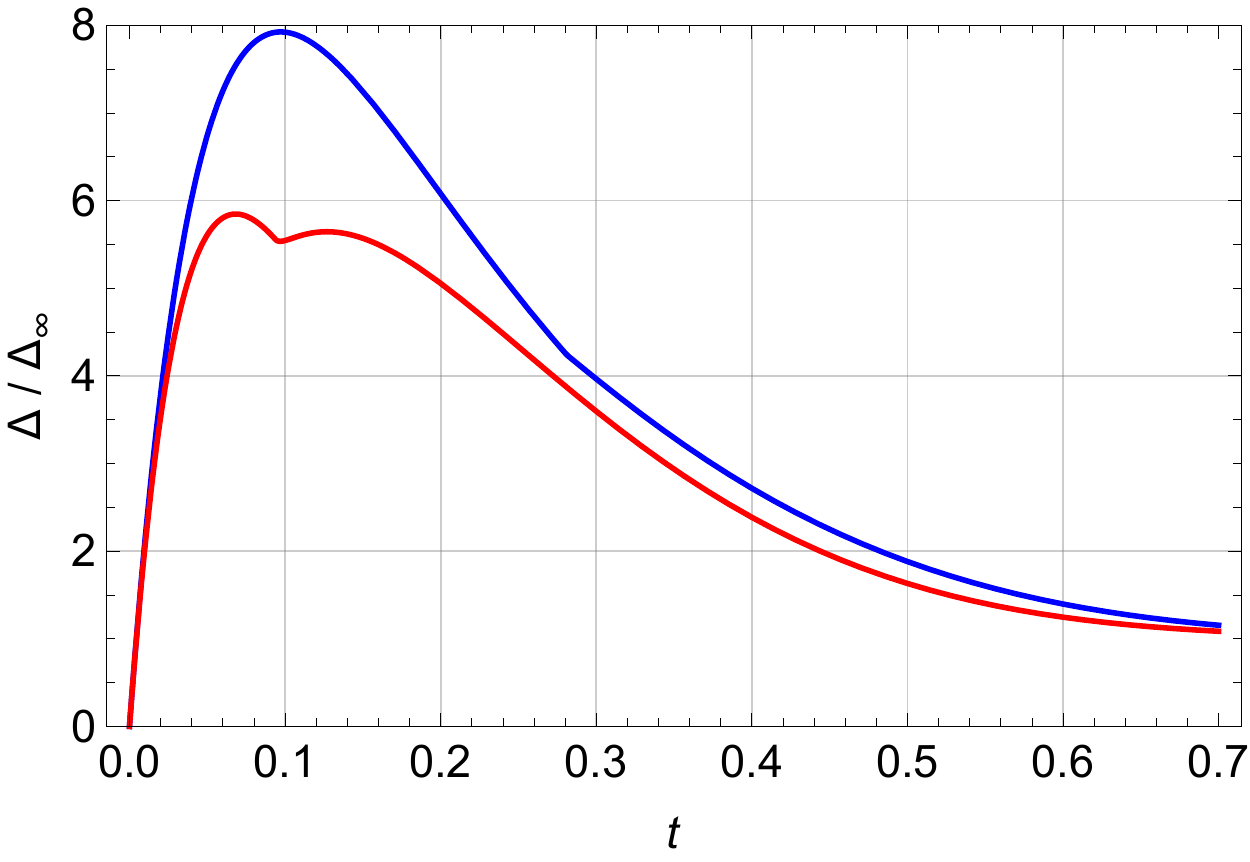}
\caption{(Color online) Normalised distance versus time $t$ (dimensionless) between the transient states for the case of the qubit initially in a maximally entangled state, blue (top) curve, and the best achievable distance for a single qubit, bottom (red) curve (i.e.\ the envelope of Fig. \ref{fig1} (Bottom)).}
\label{fig2}
\end{figure}

In Fig.\ 2 we compare this expression to the example from Fig.\ 1, and it is clear that entanglement is providing a significant advantage. Note at time $t \approx 0.28$ there is a slight kink in the blue (top) curve for $\Delta_{\phi^+}$, this occurs at the time where $C=D$, i.e. the time that the two trajectories crossed for a qubit beginning in the excited state $\theta=0$. This makes sense if one thinks of the maximally mixed reduced state of the qubit as a mixture of ground and excited states. 

In conclusion, we have analysed in detail a simple example of the general and abstract procedure of determining how well can we operationally distinguish two completely positive maps $\mathcal E_1$ and $\mathcal E_2$. The general metric this induces is known as the diamond norm \cite{KitaevShenBook}. While the abstract formulation has great power and sweeping applicability, we have seen that analysing the details of the rich dynamics induced by even one of the simplest maps - generalised amplitude damping - can provide interesting physical insights to as simple and fundamental a process as thermal equilibration of a single qubit.  In particular, we find that allowing the qubit to thermalise is sub-optimal.  Furthermore, exploiting both quantum coherence and entanglement enhances the performance of the single-qubit thermometer. 

\begin{acknowledgments}
The authors would like to thank Mark Mitchison for his helpful comments and Jian-Feng Kong for early contributions to this project.
SJ is funded by EPSRC grant
EP/K022512/1. TR supported by the Leverhulme Trust. TMS was supported by the Australian Research Council via the Centre of Excellence in Engineered Quantum Systems (EQuS) CE110001013. 
\end{acknowledgments}

\end{document}